\documentclass[11pt]{article}
\usepackage{amssymb}
\usepackage{axodraw}

\newcommand{\bi}{\bibitem}
\newcommand{\be}{\begin{eqnarray}}
\newcommand{\ee}{\end{eqnarray}}
\newcommand{\rar}{\rightarrow}

\begin{document}

\title{Rare Decays in Theories with LGS\footnote{Talk given at the 
{\it International Workshop On The Search For 
Baryon and Lepton Number Violation}, 
20~--~22~September~2007, Berkeley, California, USA.}} 

\author{Cosimo Bambi}

\date{}

\maketitle

\vspace{-1.0cm}

\begin{center}
Department of Physics and Astronomy \\ 
Wayne State University \\
Detroit, MI 48201, USA
\end{center}

\vspace{0.5cm}

\begin{abstract}
Gravitational decays in particle physics are expected
to violate the B and L quantum numbers. In the standard 
theory, where gravity is weak because of the huge value 
of Planck mass, the observation of the phenomenon is far 
from present and probably future experimental sensitivity. 
On the other hand, in theories with low gravity scale, 
where gravity becomes stronger at shorter distances, these 
processes may be dangerous, predicting a too short proton 
lifetime. Here I discuss a possible picture of gravitational 
decays, which is consistent with present experimental 
bounds for a true gravity scale of few TeV and suggests 
the possibility of observing B and L violating decays 
with a minor improvement of the present experimental 
capability.
\end{abstract}

\vspace{1cm}

In this talk I review the picture of gravitational
decays mediated by black holes I proposed in ref.~\cite{bambi1}
with Alexander Dolgov and Katherine Freese. The model
is quite speculative, but predicts B and L violating 
processes close to the existing 
experimental bounds for a fundamental gravity scale 
of few TeV. The model could also have relevant implications 
in cosmology, where B violating decays are necessary to
explain the observed matter-antimatter asymmetry in our 
universe~\cite{bambi2}.

\section{Gravitational Decays}

There are good arguments to believe that classical Black
Holes (BHs) violate global charges such as the Baryonic (B) 
and the Leptonic (L) quantum numbers~\cite{stoj}. So, if we 
include gravitational interactions in particle physics, 
we can expect the possibility of B and L violating decays
mediated by tiny BHs. The idea was indeed put forward in
1976 by Zeldovich~\cite{zeldo} and a rough estimate of the 
proton lifetime is the following. We take the proton as a 
box of side equal to its Compton wavelength $\lambda_p \sim 1/m_p$ 
with three point-like quarks inside and we consider the reaction
\be\label{qq-barql}
q + q \rar \bar{q} + l \, ,
\ee
where $q$ is a quark, $\bar q$ an anti-quark and $l$ a charged 
lepton. Here the two quarks collide and form a virtual BH, which
lastly decays violating global charges but conserving energy, 
angular momentum and electric charge. The rate of the 
process~(\ref{qq-barql}), $\Gamma_p$, is by definition
\be\label{dotn-n}
\Gamma_p = \dot{n} / n = n \, \sigma_{BH} \, ,
\ee
where $n \sim m_p^3$ is the quark number density inside the 
proton and $\sigma_{BH}$ is the cross-section of the B
violating process. Since the interaction arises from a 
dimension six operator, the amplitude has a factor 
$1/M_{Pl}^2$ and the cross section can be estimated as 
\be\label{sigma-BH}
\sigma_{BH} \sim m_p^2/M_{Pl}^4 \, .
\ee
Hence, the gravitational proton decay rate is
\be\label{sf-rate}
\Gamma_p \sim \frac{m_p^5}{M_{Pl}^4} \, .
\ee
Inserting the standard Planck mass $M_{Pl} \sim 10^{19}$ GeV
into eq.~(\ref{sf-rate}), we find that the proton lifetime is
of the order of $10^{45}$ yr, that is more than 10 orders 
of magnitude above present experimental bounds~\cite{pdg}.

\section{Theories with LGS}

In last few years, theories with Low Gravity Scale (LGS)
have attracted a lot of interest. The simplest example is
the ADD model~\cite{add}: motivated by string theory, the
observable universe would be a 4-dimensional brane embedded 
in a (4+$n$)-dimensional bulk, with the Standard Model 
particles confined to the brane, while gravity is allowed to 
propagate throughout the bulk. Here extra dimensions are 
compact and one finds that at large distances, i.e. much 
larger than the size of the extra dimensions, gravity is weak, 
because is controlled by the usual Planck mass $M_{Pl}$. On 
the other hand, at short distances, gravity becomes stronger, 
because it is controlled by the true gravity scale $M_*$ 
which can be as low as few TeV and therefore of the same 
order of magnitude of the electroweak scale. Indeed, the model 
was originally suggested to explain the hierarchy problem in 
high energy physics, that is the huge discrepancy between 
the Planck mass and the electroweak scale. The relation 
between $M_{Pl}$ and $M_*$ is
\begin{eqnarray}
M_{Pl}^2 \sim M_*^{2+n} R^n \, ,
\end{eqnarray}
where $R$ is the size of the extra dimensions. In this approach, 
however, the hierarchy problem is not really solved but shifted 
instead from the hierarchy in energies to a hierarchy in the 
size of the extra dimensions which are much larger than 1/TeV 
but much smaller than the 4-dimensional universe size.

If we put $M_* \sim 1$ TeV as fundamental gravity 
scale into eq.~(\ref{sf-rate}), we find a too short proton 
lifetime, at the level of $10^{-12}$ s, which is clearly 
inconsistent with what we observe. So, we have essentially two 
possibilities: $i)$ we must reject theories with LGS, 
because of the predicted proton lifetime, and we have to
require $M_* \gtrsim 10^{16}$~GeV~\cite{adams} or $ii)$ the 
probability of the formation of a BH is suppressed with respect 
to the Zeldovich picture. In what follows, I discuss the second 
and more fascinating option, reviewing the proposal of 
ref.~\cite{bambi1}.

\section{Classical BH Conjecture}

It is well known that classical BHs in 4 dimensions cannot have 
arbitrary large electric charge or angular momentum. Indeed, in
4 dimensions the horizon cannot be formed if~\cite{mtw}
\be\label{mass-limit}
\left( \frac{M_{BH}}{M_{Pl}}\right)^2 < \frac{Q^2}{2} +
\sqrt{ \frac{Q^4}{4} + J^2} \, ,
\ee
where $M_{BH}$, $Q$ and $J$ are respectively mass, electric charge 
and angular momentum of BH. So, classically, tiny BHs with a mass 
much smaller than the Planck mass must be electrically neutral and 
spinless.

In ref.~\cite{bambi1} I conjectured that something similar may hold
for BHs mediating gravitational decays as well. So, I suggested
that the formation of an intermediate BH is somehow a classical
process: the event horizon is formed only in particle collisions
(i.e. in the $s$-channel of a reaction, not in the $t$-channel) and
out of positive energies (i.e. time-energy uncertainty relation 
cannot create a BH with mass larger than the energy of the initial
particle(s)). This implies that the decay of particles much lighter 
than the fundamental gravity scale can be mediated only by BH
devoid of any quantum number. Of course, such a condition suppresses
the process and, as I show in the next section, we can have
a true gravity scale as low as few TeV without contradiction with
experiments. The whole picture may look very strange, but virtual
BHs are not well defined objects and it is quite probable that
the standard rules of quantum field theory are not applicable to
gravity. In absence of a quantum theory of gravity, this is the 
simplest possibility, with the advantage that its predictions are 
numerous and close to existing bounds.

Even if we cannot reliably calculate the decay rates of these 
processes, we assume they can be evaluated on dimensional
grounds, with numerical coefficients of order unity. So, we
guess that the coupling constant of BH to two fermions is
\be
g_2 \sim R_S \, E \, ,
\ee
where $E$ is the energy of all the colliding particles which make
the BH in their center of mass system, that is $E = M_{BH}$,
and $R_S$ is the BH Schwarzschild gravitational radius. On the
other hand, the creation of a BH in a multi-particle collision
should be further suppressed, because the particles must meet
in the same small volume. By dimensional arguments we can expect
that the coupling constant of BH to four fermions is
\be
g_4 \sim R_S^4 \, E \, .
\ee 
This choice of $g_4$ leads to the reasonable result that
in a 4-body collision the probability of BH creation is suppressed
by an additional small ratio square of BH volume to the interaction 
volume with respect a 2-body collision.

\section{Phenomenology}

\subsection{Leptonic and Semi-Leptonic Decays}

Let us start with the muon decay $\mu^- \rar e^- e^+ e^-$. 
In our picture, first the muon emits a virtual photon, then the 
photon produces an $e^+ e^-$ pair. Next the muon and the positron 
form a BH devoid of any quantum number. Since the BH does not
respect the family lepton number conservation, it can decay into
an $e^+ e^-$ pair, see fig.~\ref{f-m}$a$. Using the coupling
constant of BH to two fermions, we find the decay width
\be
\Gamma(\mu \rar 3e) = \frac{\alpha^2 m_\mu}{2^{11}\pi^5}\,
\left(\ln \frac{M_*^2}{m_\mu^2}\right)^2\,
\left(\frac{m_\mu}{M_*}\right)^{4(1+\frac{1}{n+1})}
\kappa^{\frac{2}{n+1}} \, ,
\ee
where $\kappa = 0.3 - 0.5$. In the case of the ordinary 
(3+1)-dimensional gravity with $M_* = M_{Pl}$, the decay rate is 
negligibly small. On the other hand, if large extra dimensions
exist, the decay could be on the verge of the experimental
discovery: the present experimental constraint is~\cite{pdg}
\begin{eqnarray}
BR(\mu^- \rightarrow e^-e^+e^-)\,\Big|_{Exp}
< 1.0 \cdot 10^{-12} 
\end{eqnarray} 
and requires that $M_*$ is not smaller than 1 -- 10 TeV (the exact
value depends on the number of large extra dimensions $n$), that 
is around the range where $M_*$ should be in order to 
explain the hierarchy problem.

Other promising and interesting reactions involving electrons and muons
are the processes $e^+ + e^- \rar \mu+ e$ and $\mu \rar e \gamma$
(see fig.~\ref{f-m}$b$), 
where the predicted cross-section and branching ratio for 
$M_* \sim$ few TeV are surprisingly close to the current limits.

Of course, we can also consider tau decays. However,
even if the tau lepton is heavier than the muon, its lifetime is shorter
and the bounds on B and L violating branching ratios weaker.
So, for $M_* \sim 1$ TeV and $n = 2$ one finds that the
expected branching ratios of the decays $\tau\rar 3l$ and 
$\tau\rar l\gamma$ are around $10^{-11}$, which surely do not
contradict the existing bounds, of order of 
$10^{-6}-10^{-7}$~\cite{pdg}. On the other hand, $\tau$ decays 
with non-conservation of B and L numbers, as e.g.
$\tau^- \rar e^-e^+ \bar p$, $e^-e^-p$ and analogous ones with
neutrons and neutrinos, would be strongly suppressed, because
here the BH emits three quarks and one lepton (instead of two 
leptons): as in BH creation, multi-particle decay is
suppressed due to the necessity for several particles to meet
in the same small volume.

\begin{center}
\begin{figure}
\begin{picture}(340,85)(-5,0)
\Text(48,80)[]{$1a$}
\ArrowLine(40,25)(80,25)
\ArrowLine(80,25)(130,25)
\PhotonArc(110,25)(30,90,180){3}{5}
\ArrowLine(110,55)(180,75)
\ArrowLine(110,55)(130,25)
\ArrowLine(150,25)(180,45)
\ArrowLine(150,25)(180,5)
\GCirc(140,25){10}{0.5}
\Text(62,15)[]{$\mu^-$}
\Text(108,15)[]{$\mu^-$}
\Text(80,55)[]{$\gamma$}
\Text(132,47)[]{$e^+$}
\Vertex(80,25){2}
\Vertex(110,55){2}
\Vertex(130,25){2}
\Vertex(150,25){2}
\Text(135,5)[]{$BH$}
\Text(147,77)[]{$e^-$}
\Text(175,52)[]{$e^+$}
\Text(175,23)[]{$e^-$}
\Text(226,80)[]{$1b$}
\ArrowLine(220,25)(260,25)
\ArrowLine(260,25)(310,25)
\PhotonArc(290,25)(30,90,180){3}{5}
\ArrowLine(290,55)(350,55)
\ArrowLine(290,55)(310,25)
\ArrowLine(330,25)(350,55)
\ArrowLine(330,25)(380,15)
\Photon(350,55)(380,75){3}{4}
\GCirc(320,25){10}{0.5}
\Text(244,15)[]{$\mu^-$}
\Text(289,15)[]{$\mu^-$}
\Text(260,55)[]{$\gamma$}
\Text(314,42)[]{$e^+$}
\Vertex(260,25){2}
\Vertex(290,55){2}
\Vertex(310,25){2}
\Vertex(330,25){2}
\Vertex(350,55){2}
\Text(315,5)[]{$BH$}
\Text(325,68)[]{$e^-$}
\Text(354,42)[]{$e^+$}
\Text(360,10)[]{$e^-$}
\Text(375,58)[]{$\gamma$}
\end{picture}
\caption{$a)$ Muon decay $\mu \rightarrow 3e$. 
 $b)$ Muon decay $\mu \rightarrow e \gamma$.}
\label{f-m}
\end{figure}
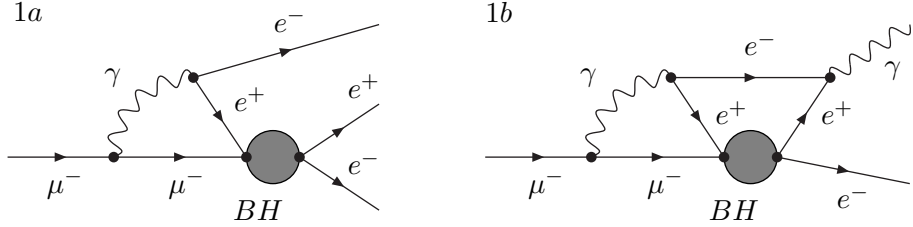
\end{center}

\subsection{$K$-meson Decays}

Good candidates for looking for L number violations are rare
decays of neutral and charged $K$-mesons. Let us focus on the
$K^0$-meson: if two quarks constituting $K^0$-meson might form 
BH, this BH could decay into any neutral combination of two 
leptons, $e^+e^-, \mu^+\mu^-$ and $\mu^\pm e^\mp$. It is easy
to see that we do not contradict experimental bounds if we
take $M_* > 3 \, (4)$ TeV for $n = 2 \, (7)$. However, it
would be natural to expect that BH has the quantum numbers of 
the vacuum, i.e. it is a scalar object. Hence the $K$-meson, 
which is a pseudoscalar, cannot transform to BH directly,
but should emit some other particle in such a way that the 
remaining combination of the quark-antiquark system would be 
scalar. The simplest way is to emit a $\pi^0$-meson,
while the remainder would make a BH which would decay into 
$l\bar l$, see fig.~\ref{f-k}. The lifetime of the decay 
$K \rar \pi ll$ is equal to
\be
\tau(K \rar \pi ll ) &=& 0.85 \cdot 10^2\,{\rm s}\, 
\left(g_{K\pi S} m_\pi \right)^{-2}\,
\left(\frac{M_*}{{\rm TeV}}\right)^{4 + \frac{4}{n+1}}
\left(\frac{{\rm TeV}}{m_K}\right)^{\frac{4}{n+1}-\frac{4}{3}}
\cdot \nonumber\\ && \quad \cdot
\left(\frac{300\,{\rm MeV}}{m_q}\right)^4 \,
\frac{6.4 \cdot 10^{-3}}{f_n} \, ,
\ee
where $g_{K\pi S} $ is the coupling constant of $K$ and $\pi$ 
to the scalar state of quark-antiquark pair and $f_n$ is related 
to integration over phase space
\be
f_n = \int_{\mu}^{(1+\mu^2)/2} dx\,\sqrt{x^2-\mu^2}\,
\left(1+\mu^2 -2x\right)^{1+\frac{2}{n+1}} \, .
\ee 
Here $\mu = m_\pi/m_K$. The factor $6.4\cdot 10^{-3}/f_n$
is equal to 1 for $n=2$, to 0.82 for $n=3$, and to 0.58 for $n=7$.   
In any case, for $M_*$ at the level of few TeV we still continue
to predict branching ratios close to existing bounds. 
So, if BHs are only scalar objects and parity is conserved, 
there are some interesting features/signatures: $i)$ the dominant 
anomalous decay mode is 3 body, $ii)$ the charge of the emitted pion 
is the same as the charge of the initial $K$, $iii)$ the probabilities 
of the decays with charged and neutral leptons in the final states 
are approximately the same. The rather large magnitude of the 
branching ratios of these anomalous decays of $K$-mesons
make them very interesting/promising candidates in the search
for non-conservation of global L quantum numbers.

\begin{center}
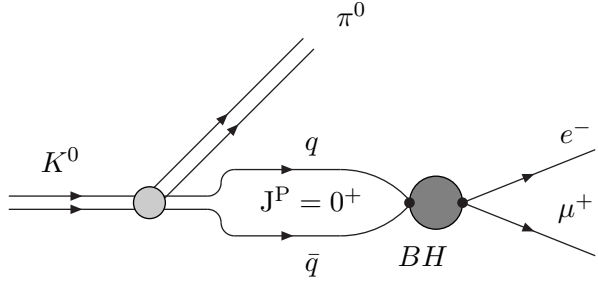
\begin{figure}
\begin{picture}(340,110)(-5,0)
\ArrowLine(100,31)(151,31)
\ArrowLine(100,26)(151,26)
\ArrowLine(151,30.5)(211,90.5)
\ArrowLine(155,26.5)(215,86.5)
\Line(155,31)(175,31)
\CArc(175,36)(5,270,360)
\CArc(185,36)(5,90,180)
\Line(155,26)(175,26)
\CArc(175,21)(5,0,90)
\CArc(185,21)(5,180,270)
\ArrowLine(185,41)(225,41)
\ArrowLine(185,16)(225,16)
\CArc(225,6)(35,40,90)
\CArc(225,51)(35,270,320)
\ArrowLine(271,28.5)(321,48.5)
\ArrowLine(271,28.5)(321,8.5)
\GCirc(261,28.5){10}{0.5}
\Text(257,8.5)[]{$BH$}
\Text(215,50)[]{$q$}
\Text(215,5)[]{${\bar q}$}
\Text(215,30)[]{${\rm J^P} = 0^+$}
\Text(230,100)[]{$\pi^0$}
\Text(315,55.5)[]{$e^-$}
\Text(315,26.5)[]{$\mu^+$}
\Text(120,45)[]{$K^0$}
\GCirc(153,28.5){6}{0.8}
\Vertex(251,28.5){2}
\Vertex(271,28.5){2}
\end{picture}
\caption{Kaon decay $K^0 \rightarrow \pi^0 e^- \mu^+$.}
\label{f-k}
\end{figure}
\end{center}

\subsection{Proton Decay}

As for the proton decay, now we need a 4-body collision in order
to create an electrically neutral, colorless and non-rotating
BH and the probability of the process is strongly suppressed,
see fig.~\ref{f-p}. Indeed one finds that the lifetime of the 
proton with respect to the inclusive decay $p\rar l^+ l^- l^+$ is
\be
\tau_p \approx 10^{29}\,{\rm yr}\,
\left(\frac{M_*}{{\rm  TeV}}\right)^{10+\frac{10}{n+1}}
\left(\frac{{\rm  TeV}}{m_p}\right)^{\frac{10}{n+1}-\frac{10}{3}}\,
\left(\frac{100{\rm MeV}}{\Lambda}\right)^{6}\,
\ln^{-2}\left(M_*/\rm TeV\right)\, f^{-1}_p(n) \, ,
\ee
where $\Lambda \sim 100$~MeV is the QCD scale (basically the 
inverse proton size) and $f_p(n) $ a numerical factor equal to 1, 
1.3 and 2.2 for $n =2, 3$ and 7 respectively. The best experimental 
lower bounds, at the level of $\tau_p >10^{33}$~yr~\cite{pdg}, are 
established for the modes $p\rar e^+\pi^0$ and $p\rar \nu K^+$. 
For all other 2-body and some 3-body modes the bounds are at the level 
of $10^{32}$~yr. So, if we believe that a BH cannot go into a 
pseudoscalar particle, the dominant decay modes are $p \rar l^+l^+l^-$ 
($l=e,\mu$). The experimental bounds, at the level 
$(5 - 8) \cdot 10^{32}$ yr, are consistent with the theoretical
model if the fundamental gravity scale $M_*$ is slightly larger than 
2 (8) TeV for 2 (7) large extra dimensions.

\begin{center}
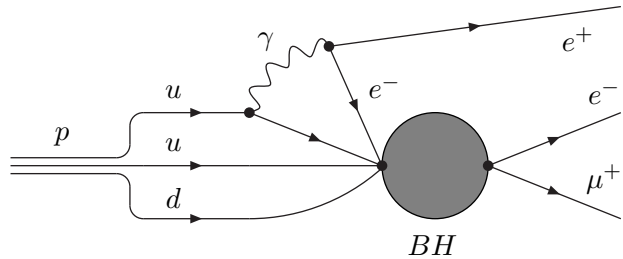
\begin{figure}
\begin{picture}(340,110)(-5,-20)
\Line(95,18)(135,18)
\CArc(135,23)(5,270,360)
\Line(140,23)(140,30)
\CArc(145,30)(5,90,180)
\Line(95,15)(145,15)
\Line(95,12)(135,12)
\CArc(135,7)(5,0,90)
\Line(140,7)(140,0)
\CArc(145,0)(5,180,270)
\ArrowLine(145,35)(185,35)
\ArrowLine(185,35)(235,15)
\ArrowLine(145,15)(185,15)
\Line(185,15)(235,15)
\ArrowLine(145,-5)(185,-5)
\CArc(185,70)(75,270,315)
\PhotonArc(215,30)(30,90,170){3}{4}
\ArrowLine(215,60)(325,75)
\ArrowLine(215,60)(235,15)
\ArrowLine(275,15)(325,35)
\ArrowLine(275,15)(325,-5)
\GCirc(255,15){20}{0.5}
\Text(115,26)[]{$p$}
\Text(157,43)[]{$u$}
\Text(237,45)[]{$e^-$}
\Text(157,23)[]{$u$}
\Text(157,3)[]{$d$}
\Text(192,62)[]{$\gamma$}
\Vertex(235,15){2}
\Vertex(275,15){2}
\Vertex(215,60){2}
\Vertex(185,35){2}
\Text(255,-15)[]{$BH$}
\Text(310,63)[]{$e^+$}
\Text(320,43)[]{$e^-$}
\Text(320,10)[]{$\mu^+$}
\end{picture}
\caption{Proton decay.}
\label{f-p}
\end{figure}
\end{center}

\subsection{$n - \bar{n}$ Oscillation}

A process where non-conservation of baryons is actively
studied by experiments is neutron-antineutron transformation. 
In the framework of the approach presented in 
ref.~\cite{bambi1}, the $n - \bar n$ oscillations are described 
by the diagram of fig.~\ref{f-n1}. A rough estimate of
the time of neutron-antineutron oscillations is
\be\label{tau-nn}
\tau_{n\bar n}= \left[\frac{2\alpha}{\pi}\ln 
\left(\frac{M_*}{m_Z}\right)\right]^{-2} 
\left(\frac{M_*}{\Lambda}\right)^{7+\frac{8}{n+1}}
\,\Lambda^{-1} \, . 
\ee
For example, taking $n=2$, $M_* \sim 1$ TeV and $\Lambda = 100$ MeV
the oscillation time is about $3\cdot 10^{19}$ s, that is 12 -- 13
orders of magnitude above the existing experimental limit~\cite{pdg}: 
direct searches for $n\rar\bar{n}$ processes using reactor neutrons 
put the upper limit $\tau_{n\bar{n}} > 8.6 \cdot 10^7$ s on the 
mean time of transition in vacuum, while the limit found from nuclei 
stability is slightly stronger, $\tau_{n\bar{n}} > 1.3 \cdot 10^8$.
If the theoretical prediction of eq.~(\ref{tau-nn}) were true, 
the chances to observe $(n-\bar n)$-oscillations in the reasonable 
future are negligible.

One can obtain much more optimistic predictions if there exist 
supersymmetric partners of the usual particles. In this case, 
one of the quarks in the neutron can emit a neutralino, $\chi^0$, 
and become a squark, $\tilde q$. This $\tilde q$, together with 
remaining quarks, can form a neutral and spinless BH. This BH in 
turn may decay into two antiquarks, $2\bar q$, and anti-squark, 
$\bar{\tilde q}$. The latter captures $\chi^0$ and becomes 
the usual antiquark, $\bar q$. This completes the transformation 
of three quarks into three antiquarks (see fig.~\ref{f-n2}).
The estimated time of $(n-\bar n)$-oscillations would be 
\be
\tau_{n\bar n} \approx 3\cdot 10^9\,
{\rm sec}\,
\cdot 10^{\frac{12}{n+1} -4}\,
\left(\frac{100 \,{\rm MeV }}{\Lambda}\right)^6\,
\left(\frac{m_{SUSY}}{300{\rm GeV }}\right)\,
\left(\frac{{\rm GeV }}{M_{BH}}\right)^{\frac{4}{n+1}}\,
\left(\frac{M_{*}}{{\rm TeV }}\right)^{\frac{4(n+2)}{n+1}} \, .
\ee
This result looks quite promising. If  $M_*$ is not too much larger 
than 1 TeV and the SUSY partners are not far from 300 GeV, the 
chances to observe neutron-antineutron transformations are very good.
On the other hand, the contribution of SUSY partners to proton decay
is negligible.

\begin{center}
\begin{figure}
\begin{picture}(340,110)(-5,-20)
\Line(50,18)(90,18)
\CArc(90,23)(5,270,360)
\Line(95,23)(95,30)
\CArc(100,30)(5,90,180)
\Line(50,15)(100,15)
\Line(50,12)(90,12)
\CArc(90,7)(5,0,90)
\Line(95,7)(95,0)
\CArc(100,0)(5,180,270)
\ArrowLine(100,35)(140,35)
\ArrowLine(140,35)(190,15)
\ArrowLine(100,15)(140,15)
\Line(140,15)(190,15)
\ArrowLine(100,-5)(140,-5)
\CArc(140,70)(75,270,315)
\PhotonArc(210,-20)(90,118,142){3}{4}
\PhotonArc(210,-20)(90,38,62){3}{4}
\ArrowArcn(210,-20)(90,118,62)
\ArrowLine(170,60)(190,15)
\ArrowLine(230,15)(250,60)
\ArrowLine(230,15)(280,35)
\ArrowLine(280,35)(320,35)
\Line(230,15)(280,15)
\ArrowLine(280,15)(320,15)
\CArc(280,70)(75,225,270)
\ArrowLine(280,-5)(320,-5)
\GCirc(210,15){20}{0.5}
\CArc(320,30)(5,0,90)
\Line(325,30)(325,23)
\CArc(330,23)(5,180,270)
\Line(330,18)(370,18)
\Line(320,15)(370,15)
\CArc(320,0)(5,270,360)
\Line(325,0)(325,7)
\CArc(330,7)(5,90,180)
\Line(330,12)(370,12)
\Text(70,26)[]{$n$}
\Text(112,43)[]{$u$}
\Text(191,39)[]{$\nu_l$}
\Text(112,23)[]{$d$}
\Text(112,3)[]{$d$}
\Text(147,60)[]{$Z^0$}
\Vertex(190,15){2}
\Vertex(230,15){2}
\Vertex(170,60){2}
\Vertex(250,60){2}
\Vertex(140,35){2}
\Vertex(280,35){2}
\Text(210,-15)[]{$BH$}
\Text(212,60)[]{$\nu_l$}
\Text(231,39)[]{$\nu_l$}
\Text(310,43)[]{${\bar u}$}
\Text(310,23)[]{${\bar d}$}
\Text(310,3)[]{${\bar d}$}
\Text(279,60)[]{$Z^0$}
\Text(350,26)[]{${\bar n}$}
\end{picture}
\caption{($n - \bar n$)-oscillation.}
\label{f-n1}
\end{figure}
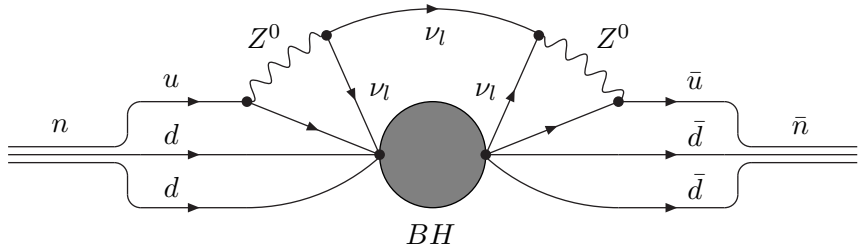
\end{center}
\begin{center}
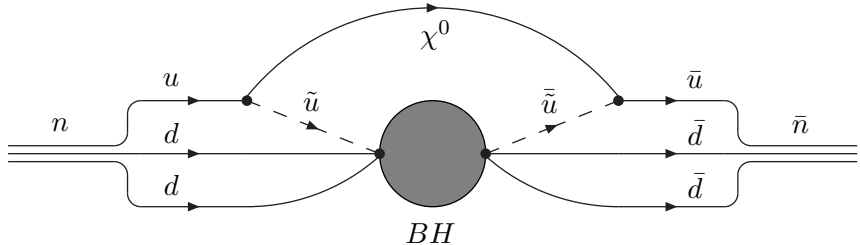
\begin{figure}
\begin{picture}(340,110)(-5,-20)
\Line(50,18)(90,18)
\CArc(90,23)(5,270,360)
\Line(95,23)(95,30)
\CArc(100,30)(5,90,180)
\Line(50,15)(100,15)
\Line(50,12)(90,12)
\CArc(90,7)(5,0,90)
\Line(95,7)(95,0)
\CArc(100,0)(5,180,270)
\ArrowLine(100,35)(140,35)
\DashArrowLine(140,35)(190,15){5}
\ArrowLine(100,15)(140,15)
\Line(140,15)(190,15)
\ArrowLine(100,-5)(140,-5)
\CArc(140,70)(75,270,315)
\ArrowArcn(210,-20)(90,142,38)
\DashArrowLine(230,15)(280,35){5}
\ArrowLine(280,35)(320,35)
\Line(230,15)(280,15)
\ArrowLine(280,15)(320,15)
\CArc(280,70)(75,225,270)
\ArrowLine(280,-5)(320,-5)
\GCirc(210,15){20}{0.5}
\CArc(320,30)(5,0,90)
\Line(325,30)(325,23)
\CArc(330,23)(5,180,270)
\Line(330,18)(370,18)
\Line(320,15)(370,15)
\CArc(320,0)(5,270,360)
\Line(325,0)(325,7)
\CArc(330,7)(5,90,180)
\Line(330,12)(370,12)
\Text(70,26)[]{$n$}
\Text(112,43)[]{$u$}
\Text(112,23)[]{$d$}
\Text(112,3)[]{$d$}
\Text(165,35)[]{$\tilde{u}$}
\Vertex(190,15){2}
\Vertex(230,15){2}
\Vertex(140,35){2}
\Vertex(280,35){2}
\Text(210,-15)[]{$BH$}
\Text(212,60)[]{$\chi^0$}
\Text(310,43)[]{${\bar u}$}
\Text(310,23)[]{${\bar d}$}
\Text(310,3)[]{${\bar d}$}
\Text(255,35)[]{$\bar{\tilde{u}}$}
\Text(350,26)[]{${\bar n}$}
\end{picture}
\caption{($n - \bar n$)-oscillation with SUSY particles.}
\label{f-n2}
\end{figure}
\end{center}

\subsection{Summary}

Table~\ref{tab} presents the most promising processes
for the observation of B and/or L number violation in the
the case of a fundamental gravity scale in TeV range.
The second column of the table reports the existing experimental 
bounds. The third column the lower bounds on the fundamental gravity
scale $M_*$ in the case of 2 (7) large extra dimensions.

\begin{table}[ht]
\begin{center}
\begin{tabular}{||c|c|c||}
\hline \hline
& & \\
{Process} & {Experiment} & {${M_*}$, ${n = 2 \, (7)}$} \\
& & \\
\hline \hline 
& & \\
{${p \rar eee}$} & {$\quad$ ${\tau > 10^{33}}$ yr $\quad$} & {$>$ 2 (8)} \\
\hline
& & \\
{${\mu \rar \gamma e}$} & {${BR < 10^{-11}}$} & $>$ 1 (10) \\
\hline
& & \\
{${\mu \rar eee}$} & {${BR < 10^{-12}}$} & {$>$ 1 (10)} \\
\hline
& & \\
{${K \rar \mu e}$} & {${BR < 10^{-12}}$} & {$>$ 3 (4)} \\
\hline
& & \\
{${K \rar \pi \mu e}$} & {${BR < 10^{-10}}$} & {$>$ 1 (1)} \\
\hline
& & \\
{${n \leftrightarrow \bar{n}}$} & {${\tau > 10^8}$ s} & {$>$ 1 (3) (MSSM)} \\
\hline \hline
\end{tabular}
\caption{\label{tab} Summary of the most promising processes. 
Fundamental gravity scale $M_*$ in TeV.}
\end{center}
\end{table}

\section*{Acknowledgments}

This work was supported in part by NSF under grant PHY-0547794 
and by DOE under contract DE-FG02-96ER41005.

\end{document}